\pdfoutput=1

\documentclass[%
 reprint,
floatfix,
 amsmath,amssymb,
 superscriptaddress,
 jcp,
 aip
]{revtex4-2}

\usepackage{siunitx}

\usepackage{xcolor}
\definecolor{ALUblue}{RGB}{0,74,153}
\definecolor{ALUwblue}{RGB}{87,129,189}
\definecolor{ALUred}{RGB}{193,0,42}
\definecolor{ALUgrau}{RGB}{154,155,156}
\definecolor{mygreen}{RGB}{0,150,0}
\definecolor{myorange}{RGB}{254, 211, 0}

\usepackage[hyperfootnotes=false]{hyperref}	 % make refs, cites, links in color
\hypersetup{
  colorlinks,
  citecolor=ALUblue,
  linkcolor=ALUblue,
  urlcolor=ALUblue
}

\usepackage[nameinlink]{cleveref}

 \crefname{equation}{{\color{black}{Eq.}}}{{\color{black}{Eqs.}}}
 \creflabelformat{equation}{#2\textup{(#1)}#3}
 \crefname{figure}{Fig.}{Figs.}
 \Crefname{figure}{Figs.}{Figs.} %% use with multiple panels
 \crefname{table}{Table}{Tables}
 \crefname{appendix}{{\color{black}{Sec.}}}{{\color{black}{Secs.}}}
 \crefname{section}{{\color{black}{Sec.}}}{{\color{black}{Secs.}}}

\usepackage{graphicx}% Include figure files
\graphicspath{{figures/}}
 
\usepackage{caption}
\usepackage{ragged2e}
\DeclareCaptionJustification{justified}{\justifying}
\let\originalleft\left
\let\originalright\right
\renewcommand{\left}{\mathopen{}\mathclose\bgroup\originalleft}
\renewcommand{\right}{\aftergroup\egroup\originalright}

\newcommand{\tx}[1]{\text{\tiny #1}}
\newcommand{\tb}[1]{_\tx{#1}}
\newcommand{\tu}[1]{^\tx{#1}}

\newcommand{\del}{\partial}
\newcommand{\delfrac}[2]{\frac{\del #1}{\del #2}}

\newcommand{\EQ}[1]{\begin{align}#1\end{align}}

\newcommand{\C}{_\tx{c}}
\newcommand{\IN}{_\tx{in}}
 \newcommand{\pan}[1]{{\color{ALUblue}#1}}

\begin{document}

\title{Feedback-controlled solute transport through chemo-responsive polymer membranes}% 

\author{Sebastian Milster}%
\affiliation{Applied Theoretical Physics – Computational Physics, Physikalisches Institut, Albert-Ludwigs-Universität Freiburg, Hermann-Herder Strasse 3,
D-79104 Freiburg, Germany}

\author{Won Kyu Kim}
\affiliation{Korea Institute for Advanced Study, Seoul 02455, Republic of Korea}%

\author{Joachim Dzubiella}
\email{joachim.dzubiella@physik.uni-freiburg.de}
\affiliation{Applied Theoretical Physics – Computational Physics, Physikalisches Institut, Albert-Ludwigs-Universität Freiburg, Hermann-Herder Strasse 3,
D-79104 Freiburg, Germany}
\affiliation{Research Group for Simulations of Energy Materials, Helmholtz-Zentrum Berlin für Materialien und Energie, Hahn-Meitner-Platz 1, D-14109
Berlin, Germany}

\date{\today}

\begin{abstract}

%%%

{Polymer membranes are typically assumed to be inert and nonresponsive to the flux and density of the permeating particles in transport processes. Here, we study theoretically the consequences of membrane responsiveness and feedback on the steady-state  force--flux relations and membrane permeability using a nonlinear-feedback solution-diffusion model of transport through a slab-like membrane.  Therein, the solute concentration inside the membrane depends on the bulk concentration, $c_0$, the driving force, $f$, and the polymer volume fraction, $\phi$. In our model, solute accumulation in the membrane causes a sigmoidal volume phase transition of the polymer, changing its permeability, which, in return, affects the membrane's solute uptake. This feedback leads to nonlinear force--flux relations, $j(f)$, which we quantify in terms of the system's differential permeability, $\mathcal{P}\tb{sys}^{\Delta}\propto {\mathrm{d}j}/{\mathrm{d}f}$. We find that the membrane feedback can increase or decrease the solute flux by orders of magnitude, triggered by a small change in the driving force, and largely tunable by attractive versus repulsive solute--membrane interactions. Moreover, controlling the input, $c_0$ and $f$, can lead to steady-state bistability of $\phi$ and hysteresis in the force--flux relations. This work advocates that the fine-tuning of the membrane's chemo-responsiveness will enhance the nonlinear transport control features, providing great potential for future (self-)regulating membrane devices. }

\end{abstract}

\maketitle

\section{Introduction}

The precise and selective control of molecular transport through membranes is of fundamental importance for various applications in industry and medicine, such as water purification,\cite{shannon:nature,geise2010water,Pendergast2011} food-processing,\cite{Cuperus1993,Kumar2013} nano-catalysis,\cite{Renggli2011a,roa2017catalyzed,Lu2011,msde2020modeling} drug-delivery,\cite{Li2016b,Brudno2015,Moncho-Jorda2020,Mitchell} and tissue engineering.\cite{stamatialis,Lee2018}  Modern membrane technology becomes increasingly inspired by responsive bio-membranes with nonlinear potential-, pressure- or flux-gated permeabilities, bistable behavior and memristive properties.\cite{Hodgkin1952,Henisch1974,Yang2006,Keener2009,TakheeLeeandYongChen,Lei2014,Hasan2019,Minoura1998} Such features allow the design of highly selective membrane devices that efficiently control molecular transport,  autonomously regulate the chemical milieu, and may act as logical operators, artificial synapses, or analogous filters for electrical or chemical signals. Moreover, the possible memristive properties create the foundation for information storage, adaptive responses to stimuli based upon past events, and neuromorphic systems. \cite{Hasan2019a,Jo2010,Prodromakis2010}

In general, such self-regulation premises a feedback mechanism controlling the transport properties in a nonlinear fashion.\cite{Frank2005,Giacomelli2012,Gernert2015,Holubec2022} In the scope of membrane applications this may arise from various system-dependent effects, such as autocatalysis, substrate or product inhibition,\cite{Katchalsky1968,Kepperbook} the interplay of voltage and hydrodynamic pressure,\cite{Teorell1951,Teorell1962} or, as highlighted in this work, the reciprocal impacts of molecular fluxes and membrane permeability.\cite{Hahn1973,Bell2021a,Zou1999,Leroux1999,Sparr2009,TimaCosta-Balogh2005,Aberg2009,Tikhonov2011}
In this regard many polymeric compounds offer great potential as they are versatile in their response to various physico-chemical stimuli and environmental conditions, such as temperature, electric field, solvent quality, etc.\cite{Stuart2010,Schattling2013,Wandera2010} For example, the polymer responds with a volume phase transition, either from a swollen to a collapsed state, or vice versa, in which the polymer volume fraction, $\phi$, may change by orders of magnitude.\cite{Tanaka1978,Tanaka1980,Hirokawa1998,Koga2001,Roiter2005,kim2017cosolute,Bell2021a}  Such a drastic change of the polymer's physical features, in turn, has substantial, nonlinear effects on the solute permeability of the membrane device. 
\begin{figure}[b]
\includegraphics{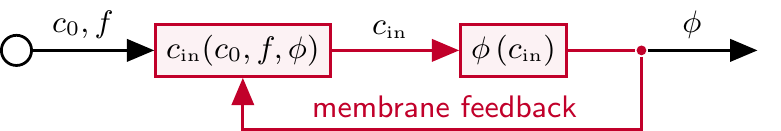}
\caption{\label{fig:mini_loop}Essential feedback loop of chemo-responsive polymer membranes pointing out the nonlinear, reciprocal dependence of the polymer volume fraction, $\phi(c\tb{in})$, and the solute concentration inside the membrane, $c\tb{in}(\phi)$. A change in the solute bulk concentration, $c_0$, or the external force, $f$, acting on the solutes has nontrivial effects on $c\tb{in}$ and $\phi$, and thus on the transport properties of the membrane.}
\end{figure}

Very illustrative examples are so-called \emph{smart gating membranes},\cite{Liu2016,Ito2002,Ito2004,bhattacharyya2012responsive,Li2020,Xie2007,Tokarev2010} which are (rather solid) porous membranes with polymer-coated channels that can reversibly open and close, triggered by external stimuli or, through autonomous feedback, by molecular recognition. Moreover, literature on the solution-diffusion model\cite{Baker,Yasuda1969,gehrke,missner2009110} suggests that the use of more flexible, responsive polymeric membranes enables feedback-controlled solute transport with further valuable features, such as multiple steady states and hysteresis transitions.\cite{Hahn1973,Siegel1995, Leroux1999, Zou1999, Bell2021a} However, more research is needed here to understand the role of the membrane feedback, especially in the presence of external driving, and how hysteresis transitions can occur.

For nonresponsive polymer membranes, we have previously shown that the Smoluchowski equation\cite{MarianvonSmoluchowski} well describes solute flux and concentration profiles under stationary nonequilibrium conditions.\cite{Kim2022} Therein, we reported that the membrane's solute uptake, $c\IN$, is not only a function of the polymer volume fraction, $\phi$, the membrane permeability, $\mathcal{P}\tb{mem}(\phi)$, and bulk concentration, $c_0$, but also tuneable in nonequilibrium by a the external driving force, $f$. The latter leads to a nonlinear flux, $j(f)$, with significant differences between the low- and high-force regimes. The nonlinear intermediate crossover was quantified using the newly introduced system's {\it differential permeability}, $\mathcal{P}\tb{sys}^{\Delta}\propto{\mathrm{d}j}/{\mathrm{d}f}$.

Motivated by the above features and open questions, in this work we turn our attention to polymer membranes that are responsive to the penetrants, and highlight the key differences compared to nonresponsive membranes. Specifically, we include a mean-field model for the polymer response in the Smoluchowski framework, i.e., $\phi\to\phi(c\IN)$ is a sigmoidal function of the average solute uptake, which enters $\mathcal{P}\tb{mem}(\phi)$ and, in turn, controls $c\IN$, leading to a membrane-intrinsic feedback mechanism [\cref{fig:mini_loop}]. Eventually, we use empirical expressions for $\mathcal{P}\tb{mem}(\phi)$ to study the feedback effect on $j$ and $\mathcal{P}\tb{sys}^{\Delta}$ as function of $c_0$ and $f$.
Compared to nonresponsive membranes, we find substantial enhancement of the nonlinear characteristics, such as an order of magnitude change in $j$ due to a very small change in $f$, and report the emergence of multiple steady sates, bifurcations, and hysteresis in the force--flux relations. 
 
\section{Theoretical framework\label{sec:smolu}}
\subsection{Steady-state Smoluchowski equation and system setup}
\begin{figure}

\begin{center}
\includegraphics{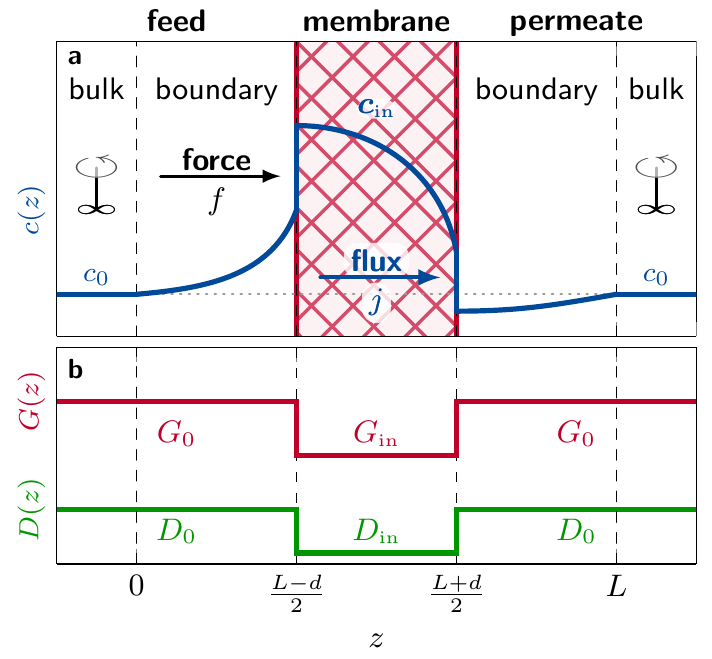}
\end{center}
\caption{\label{fig:Fick}(a): System setup showing a membrane (red) of width $d$ in $z$-direction (periodic in $x$ and $y$) in the center of the system of size $L$, and an example solute concentration profile $c(z)$ (blue) in a steady state with external driving, $f>0$. The system is in contact with identical solute bulk reservoirs with constant concentration, $c(0)=c(L)=c_0$. The concentration in the boundary and membrane layers, described by the Smoluchowski framework, is determined by $f$, and the energy and diffusion landscapes, $G(z)$ and $D(z)$, depicted in (b).}
\end{figure}

We consider the solute transport across a polymer membrane as a one-dimensional drift-diffusion process (in $z$-direction) of ideal solutes [see the system sketch in \cref{fig:Fick}\pan{(a)}]. The membrane has the width $d$, and is located in the center of the system of length $L$, yielding interfaces at $(L\pm d)/2$. It is in contact with two solute reservoirs of equal concentration $c_0$ via boundary layers on the feed and permeate sides. \cite{Grassi1999} The steady-state flux in the overdamped limit derived from the Smoluchowski equation, reads \cite{risken1996fokker}
\EQ{
j=-D(z)\left[ \delfrac{c(z)}{z} +\beta c(z)\left(\delfrac{G(z)}{z}-f\right) \right], \label{eq:smolu}
}
with the inverse temperature, $\beta=1/(k\tb{B}T)$, and the (external) driving force, $f$, which may result from various sources. We assume that the diffusion and energy landscapes, $D(z)$ and $G(z)$, are piecewise homogeneous, cf. \cref{fig:Fick}\pan{(b)}, precisely
\EQ{\renewcommand{\arraystretch}{1.25}
D(z)=\left\{\begin{array}{l l}
     D\tb{in} & \frac{L-d}{2}\le z\le\frac{L+d}{2},\\
     D_0 & \text{elsewhere},\\
            \end{array}\right. }
and
\EQ{\renewcommand{\arraystretch}{1.25}
G(z)=\left\{\begin{array}{l l}
     G\tb{in} & \frac{L-d}{2}\le z\le\frac{L+d}{2},\\
     G_0 & \text{elsewhere},\\
            \end{array}\right. }
where the subscripts `$0$' and `in' refer to the regions outside and inside the membrane, respectively.

\subsection{The membrane permeability \label{sec:Pmem}}

The quantities $D\IN$ and $\Delta G=G\IN-G_0$ define the membrane permeability in the solution--diffusion picture,\cite{Baker,Yasuda1969,gehrke,missner2009110}
\EQ{
\mathcal{P}\tb{mem}=D\tb{in}\mathcal{K}, \label{eq:P_solution_diffusion}
}
where $\mathcal{K}=\exp(-\beta \Delta G)$ is the equilibrium partitioning. The membrane permeability is a function of the polymer volume fraction, $\phi$, and depends on the solute--polymer interactions.

For not too attractive solute-polymer interactions, the solute diffusivity inside the membrane is well described by Yasuda's free-volume theory,\cite{Yasuda1969} 
\EQ{
D\tb{in}(\phi)=D_0\exp\left(-A \frac{\phi}{1-\phi}\right),\label{eq:DIN}
}
with $A$ a positive parameter accounting for steric solute--polymer effects. For dilute solute systems, the partitioning of the ideal solutes can be well approximated by\cite{Kim2020}
\EQ{
\mathcal{K}=\exp\left(-B\phi\right),\label{eq:K}
}
where $B=2 B_2v_0^{-1}$ is related to the second virial coefficient, $B_2$, rescaled by effective monomer volume $v_0$.

In fact, there exist many extended versions of scaling laws for $D\tb{in}$ and $\mathcal{K}$ which take into account further microscopic details, such as the chemistry, shape and size of the solutes, the solvent and the membrane types, and the architecture of the polymer network.\cite{Milster2019,Milster2021,Kim2020,Kanduc2018,Kanduc2021, gehrke,QuesadaPerez2021Solute,Amsden1998,Hansing2018} However, we use \cref{eq:DIN,eq:K} to explain the feedback effects of responsive polymers at the simplest level of detail.

Further, the presented framework assumes that the equilibrium findings for $ \mathcal{K}$ and $D\IN$ are also valid under moderate nonequilibrium conditions, i.e., they are independent of the flux or the force. The validity of this assumption was demonstrated in our preceding work with nonequilibrium coarse-grained simulations of membrane-bulk systems.\cite{Kim2022}

\subsection{Solute flux and concentration inside membrane}

With known $\mathcal{P}\tb{mem}$ and $c(0)=c_0$, we solve \cref{eq:smolu} to obtain the concentration profile, yielding
\EQ{
c(z)=\left[c_0-j\mathcal{I}(0,z)\right]e^{-\beta(G(z)-fz)}, \label{eq:c_general}
}
with
\EQ{
\mathcal{I}(0,z)=\int\limits_{0}^z\mathrm{d}y\frac{e^{\beta(G(y)-fy)}}{D(y)}.
}
Using $c(L)=c_0$, the flux can be expressed as\cite{Kim2022}

\EQ{
j=D_0 c_0\beta f \left[1+\left(\frac{D_0}{\mathcal{P}\tb{mem}}-1 \right)S(f)\right]^{-1}, \label{eq:j_general_easy}
}
with $S(f)=\sinh(\beta fd/2)/\sinh(\beta f L/2)$,
which determines the impact of $\mathcal{P}\tb{mem}/D_0$ in the low- and high-force limits (see Appendix \ref{sec:forcelimit} for further details).

By reinserting \cref{eq:j_general_easy}, one obtains the solute concentration profile throughout the system (see Appendix \ref{sec:profile} for full expressions). We are particularly interested in the mean solute concentration inside the membrane, $
c\tb{in}:=\langle c(z)\rangle\tb{in}=d^{-1}\int\tb{in}\mathrm{d}z\ c(z)$, with $z\tb{in}\in\left[(L-d)/2,(L+d)/2\right]$, because it is the key stimulus for our membrane response model. After integration, we obtain

 \EQ{&c\tb{in}(c_0,f,\phi)=\nonumber\\
  &c_0\mathcal{K}\ \frac{2\left({D_0}-\mathcal{P}\tb{mem}\right) S(f) \sinh \left(  {\beta f(d-L)}/{2}\right)+\beta f d {D_0} }{\beta
   fd \left[({D_0}-\mathcal{P}\tb{mem}) S(f)+\mathcal{P}\tb{mem} \right]}.\label{eq:mean_cin}
}
 
The detailed behavior of \cref{eq:mean_cin} is described in the results section with an appropriate choice of the model parameters.

\subsection{The polymer response to the solute concentration}
The above theoretical framework is straightforward for membranes with constant $\phi$. We now extend the model to membranes that are responsive (in $\phi$) to $c\IN$. Motivated by experimental, theoretical, and computational findings, \cite{Tanaka1978,Tanaka1980,Hirokawa1998,Koga2001,Roiter2005,kim2017cosolute,Bell2021a,Wolf1978,Kamemaru2018,Mohammadyarloo2022} we consider that the polymer networks undergo a sigmoidal volume phase transition in the vicinity of a crossover concentration $c\tb{c}$. We assume the following form
\EQ{ 
\phi(c\tb{in})=\phi\tb{c}\pm\frac{\Delta \phi}{2}\tanh\left(\frac{c\tb{in}-c\tb{c}}{\Delta c}\right),\label{eq:phi_eq}
}
where $\Delta\phi=(\phi\tb{max}-\phi\tb{min})$ is the maximum change, and  $\phi\tb{c}=\phi(c\tb{c})=(\phi\tb{min}+\phi\tb{max})/{2}$ the polymer volume fraction at $c\C$. Hence, we call ($c\tb{c},\phi\tb{c}$) the crossover point. The transition may occur from a swollen state ($\phi\tb{min}$) to a collapsed state ($\phi\tb{max}$) or vice versa as $c\IN$ increases (denoted by the `$\pm$'-symbol in \cref{eq:phi_eq}), depending on the interactions between the solutes, the solvent, and the polymer. Effectively attractive solute-membrane interactions ($B<0$) are expected to cause a swollen-to-collapsed transition ($+$), while a transition from the collapsed to the swollen phase ($-$) is expected for repulsive interactions ($B>0$). Further, the parameter $\Delta c$ determines the sharpness of the transition, ranging from almost irresponsive ($\Delta c\gg c\tb{c}$) to very sharp transitions ($\Delta c \ll c\tb{c}$). 

Note that \cref{eq:phi_eq} assumes continuous transitions although hysteresis has been reported in experimental studies.\cite{Koga2001,Kamemaru2018,Annaka1992,Katchalsky1972} However, this work will demonstrate that hysteresis and bistability can result from the mutual dependencies of $\phi$ and $c\tb{in}$.

Furthermore, \cref{eq:phi_eq} is a mean-field approach as it does not resolve spatial inhomogeneities of $\phi$ and $c\tb{in}$ [cf. the example concentration profile in \cref{fig:Fick}\pan{(a)}]. We assume that the system is small and that the thin membranes do not change the width in the direction of the solute flux. Despite the multiple assumptions, our simplified model enables the investigation of the effect of a responsive membrane permeability on the transport.

\begin{table}
\caption{\label{tab:parameter_summary}Summary of the model parameters and the corresponding values for $\mathcal{K}, D\IN$, and $\mathcal{P}\tb{mem}$ at $\phi\tb{min}$, $\phi\tb{max}$, and $\phi\C$. Length scales are given in units of $\sigma$, the radius of one monomer. The transition width is rescaled by the crossover concentration $c\C$ [cf. \cref{eq:phi_eq}]. Permeabilities and diffusivities are expressed in units of $D_0$, the solute bulk diffusion. The arrow ($\Rightarrow$) indicates that the presented values are direct consequences of the parameter choice. The approximate Lennard-Jones energy $\varepsilon\tb{LJ}$ stems from a comparison with the second virial coefficient, $B=2B_2v_0^{-1}$. The symbols `$+$' and `$-$' correspond the swollen-to-collapsed and the collapsed-to-swollen transition, respectively [see also \cref{eq:phi_eq}].}
\scalebox{0.89}{
 \begin{tabular}{r l S S S}
   \hline
  \hline
 \multicolumn{5}{c}{\sffamily polymer response [\cref{eq:phi_eq}]}\\
   \hline\\[-0.5em]
    &$\phi\tb{min}$ & 0.05 & \multicolumn{2}{c}{\sffamily swollen} \\
    &$\phi\tb{max}$ & 0.35& \multicolumn{2}{c}{\sffamily collapsed} \\
    \\[-0.8em]
    $\Rightarrow$&$\phi\C$ & 0.2\\
    $\Rightarrow$&$\Delta\phi$ & 0.3\\
    \ \\
      &  & { \   \sffamily sharp \ \ }& { \ \ \sffamily gradual \ \ } & {  \sffamily weak \ \  }\\\cline{3-5} 
    \\[-0.8em]
    &$\Delta c /c\C$ & 0.1 & 1.0 & 10.0 \\
    \\
    \multicolumn{5}{c}{\sffamily lengths (\cref{fig:Fick})}\\
    \hline\\[-0.5em]
    &$L/\sigma$ & 100 & \multicolumn{2}{c}{\sffamily system size} \\
    &$d/\sigma$ & 90 & \multicolumn{2}{c}{\sffamily membrane width} \\
    \ \\    
    \multicolumn{5}{c}{\sffamily solute diffusion inside membrane [\cref{eq:DIN}]}\\
    \hline\\[-0.5em]
    &$A$ & 5 \\
    \ \\[-0.8em]
    $\Rightarrow$&${D\IN(\phi\tb{min})}/{D_0}$ & 0.77 \\
    $\Rightarrow$&${D\IN(\phi\tb{c})}/{D_0}$ & 0.27 \\
    $\Rightarrow$&${D\IN(\phi\tb{max})}/{D_0}$ & 0.07\\ 
\ \\
\multicolumn{5}{c}{\sffamily solute--membrane interactions and partitioning [\cref{eq:K}]}\\
    \hline\\[-0.5em]
    &  & { \   \sffamily repulsive \ \ }& { \ \ \sffamily weakly attr. \ \ } &
    { \ \ \sffamily attractive \  }\\\cline{3-5}    
    \\[-0.8em]
  &$B$ & 5.26 &  -6.25 & -17.8 \\
  \ \\[-0.8em]
  $\Rightarrow$&$\beta\varepsilon\tb{LJ}$\ {\footnotesize \sffamily (approx.) }& 0.03  &  0.55  & 0.9  \\
      \\[-0.8em]
  $\Rightarrow$&${\mathcal{K}(\phi\tb{min})}$ & 0.77 & 1.37&2.4\\
  $\Rightarrow$&${\mathcal{K}(\phi\tb{c})}$ & 0.35  & 3.49&34.9\\
  $\Rightarrow$&${\mathcal{K}(\phi\tb{max})}$ & 0.16 & 8.91&501.2\\
      \\[-0.8em]
  $\Rightarrow$ & {\footnotesize \sffamily transition [\cref{eq:phi_eq}] }& {$-$} & {$+$} & {$+$}\\
  \ \\
  \multicolumn{5}{c}{\sffamily membrane permeability [\cref{eq:P_solution_diffusion}]}\\
    \hline\\[-0.5em]
    &  & { \   \sffamily repulsive \ \ }& { \ \ \sffamily neutral \ \ } & { \ \ \sffamily attractive \  }\\\cline{3-5}    
    \\[-0.8em]
  $\Rightarrow$&${\mathcal{P}\tb{mem}(\phi\tb{min})}/{D_0}$ & 0.59 &  1.05 & 1.9 \\
  $\Rightarrow$&${\mathcal{P}\tb{mem}(\phi\tb{c})}/{D_0}$ & 0.10 & 1.00 & 10.0 \\
  $\Rightarrow$&${\mathcal{P}\tb{mem}(\phi\tb{max})}/{D_0}$ & 0.01 &  0.60 &  33.9  \\ 
  \hline
  \hline
 \end{tabular}
 }%scalebox
\end{table}

\subsection{Model parameters}
All length scales are expressed in units of $\sigma$, the effective diameter of one monomer. We set the system size  to $L=100\sigma$ and fix the membrane width to $d=90\sigma$, i.e., the boundary layers between the membrane and the two bulk reservoirs with concentration $c_0$ have the width $5\sigma$. The concentrations $c_0$,  $c\tb{in}$ and the transition width $\Delta c$ are rescaled by the crossover concentration $c\tb{c}$ of the volume phase transition [\cref{eq:phi_eq}]. We choose three different transition widths, $\Delta c/c\tb{c}\in\left\{0.1, 1.0, 10\right\}$. The force, $\beta f \sigma$, is rescaled by the thermal energy and the solute size. The solute diffusivity inside the membrane, $D\tb{in}$, and the permeability, $\mathcal{P}$, are expressed in units of the solute bulk diffusivity $D_0$. The parameters $A$ and $B$, which enter $D\tb{in}(\phi)$ [\cref{eq:DIN}] and  $\mathcal{K}(\phi)$ [\cref{eq:K}], respectively, as well as the limits of the polymer volume fraction, $\phi\tb{min}=0.05$ and $\phi\tb{max}=0.35$, are based on our group's coarse-grained simulations.\cite{Kim2020,Milster2021} We fix $A=5$, assuming that the diffusion is dominated by steric exclusion. The interaction parameter $B$ is chosen in a way to yield three different values for the equilibrium membrane permeability at $\phi\tb{c}=0.2$, and, hence, we denote the membranes as repulsive ($\mathcal{P}\tb{mem}(\phi\tb{c})/D_0=0.1$), neutral ($\mathcal{P}\tb{mem}(\phi\tb{c})/D_0=1.0$), and attractive ($\mathcal{P}\tb{mem}(\phi\tb{c})/D_0=10.0$).  In fact, due to typical cancelling effects of $\mathcal{K}(\phi)$ and $D\tb{in}(\phi)$,\cite{Kim2020,Milster2021} we can safely assume that the permeability of our neutral membrane does not significantly deviate from unity throughout the range of $\phi$. In a system with Lennard-Jones (LJ) interactions between the solutes and the membrane monomers of equal size, the characteristic LJ interactions strengths would take the approximate values $\beta \varepsilon\approx0.03$ (repulsive), $\beta \varepsilon\approx0.55$ (weakly attractive), and $\beta \varepsilon\approx0.9$ (attractive), respectively. All parameter values and relevant quantities are summarized in \cref{tab:parameter_summary}.

\section{Results and Discussion}

\begin{figure*}
\includegraphics{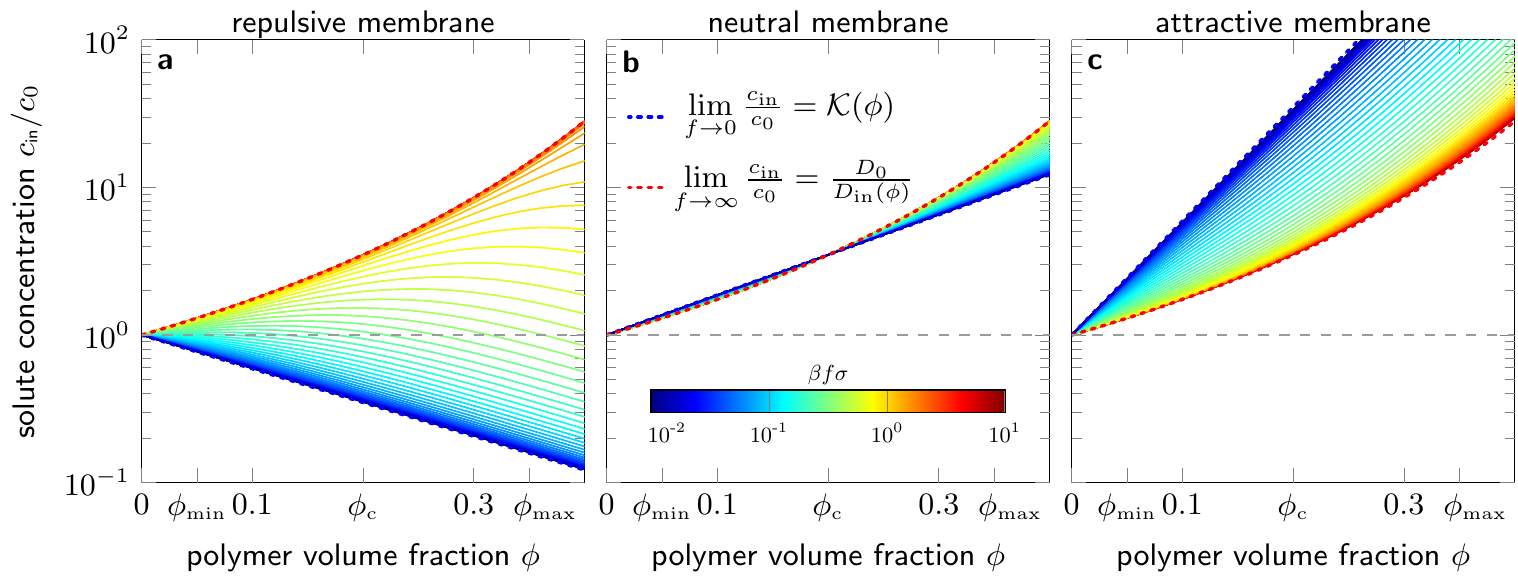} \ \ \ \
\caption{\label{fig:cin_of_phi_and_f} Mean solute concentration inside the membrane [\cref{eq:mean_cin}] as a function of $\phi$ for different values of the driving force, $f$ [color-coded, see colorbar in panel (b)], and different interaction strengths, $B\in\left\{5.26, -6.25, -17.8\right\}$, which correspond to a repulsive [(a): $\mathcal{P}_\tx{mem}(\phi_\tx{c})=0.1 D_0$], neutral [(b): $\mathcal{P}_\tx{mem}(\phi_\tx{c})=1.0 D_0$], and attractive membrane [(c): $\mathcal{P}_\tx{mem}(\phi_\tx{c})=10 D_0$], respectively, as indicated above the panels. The blue and the red dotted line represent the zero and the infinite force limits, $\lim_{f\to0}c_\tx{in}/c_0=\mathcal{K}(\phi)$ [\cref{eq:K}] and  $\lim_{f\to\infty}c_\tx{in}/c_0=D_0/D_\tx{in}(\phi)$ [\cref{eq:DIN}],  respectively. While for the repulsive membrane [panel (a)] the $c_\tx{in}$ increases with an increase in force,  it decreases for the case of the attractive membranes [panel (c)]. The solute concentration in the neutral membrane [panel (b)] depends on $f$, yet is essentially a function of $\phi$. }
\end{figure*}

\subsection{Force-controlled solute uptake}
From \cref{eq:mean_cin}, the low- and high-force limits for the mean solute concentration inside the membrane, $c\IN$, can be deduced, which has been discussed and substantiated with concentration profiles from theory and coarse-grained simulations in our previous work.\cite{Kim2022}  Here, we recapture the main findings and discuss the results for the parameters used in this work. 

In \cref{fig:cin_of_phi_and_f}, we depict $c\tb{in}$ [\cref{eq:mean_cin}] as a function of $\phi$ for different values of $\beta f\sigma\in\left[0.01,10\right]$ (color-coded) and for three different values of $\mathcal{P}\tb{mem}(\phi_c)/D_0 \in\left\{0.1,1.0,10.0\right\}$ (different panels). In the zero force limit, $c\tb{in}$  reduces to the expected equilibrium value, $\lim_{f\to0}c\tb{in}=c_0\mathcal{K}(\phi)$, which monotonously decreases for the repulsive membrane [panel (a)] and increases otherwise [panels (b) and (c)]. The same limiting result is obtained, if $\mathcal{P}\tb{mem}(\phi)=D_0\ \forall\phi$, which applies approximately for the `neutral' membrane [panel (b)]. In the high-force limit, the concentration profiles become piecewise constant with $\lim_{f\to\infty}c\tb{in}=c_0 D_0 D\tb{in}^{-1}(\phi)$,  and one further finds $\lim_{f\to\infty}j=c_0\beta f D_0 = c\tb{in}\beta f D\tb{in}$ for all membrane types, since it is independent of $\mathcal{K}$.

The solute uptake of the repulsive membrane at fixed $\phi$ increases with $f$, while it decreases in the attractive membrane [see \Cref{fig:cin_of_phi_and_f}\pan{(a)} and \labelcref{fig:cin_of_phi_and_f}\pan{(c)
}]. For `neutral' membrane [panel (b)], $c\IN$ shows no significant force dependence.

\begin{figure*}
\includegraphics{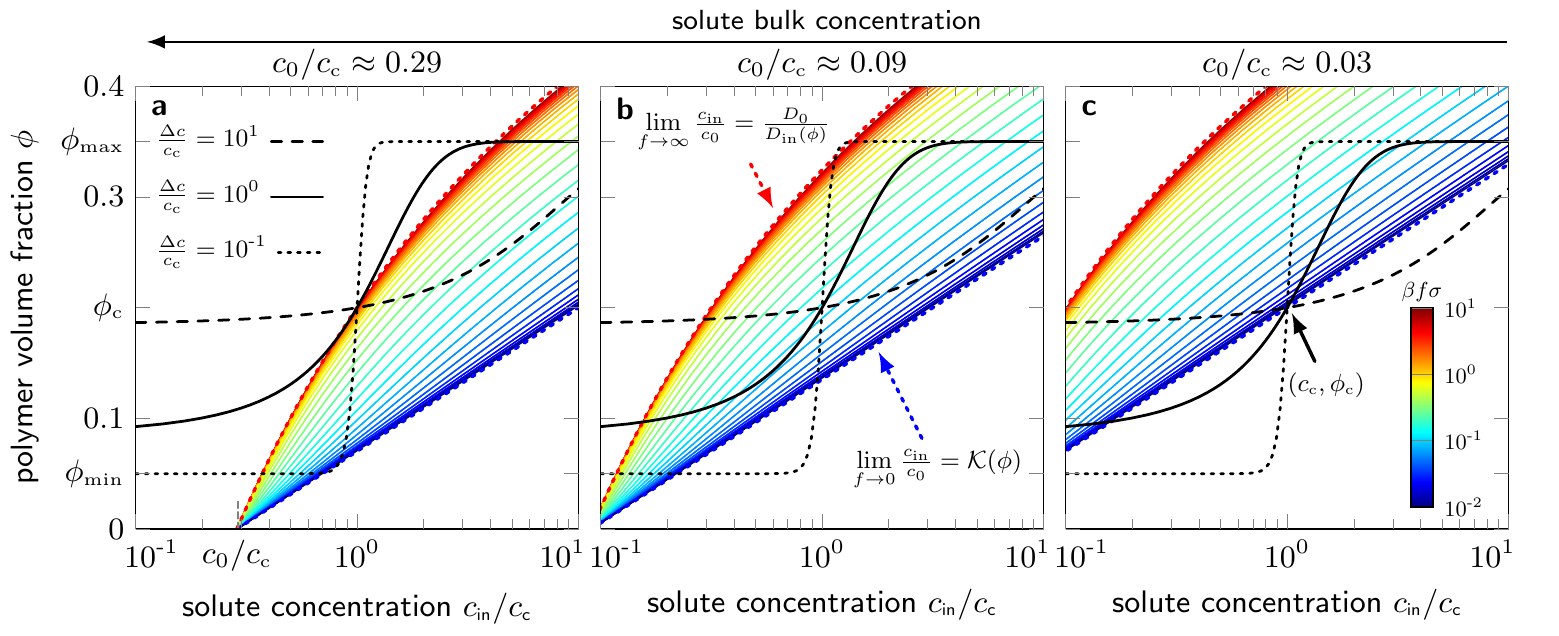}
\caption{Phase plane showing $\phi(c\IN)$ [\cref{eq:phi_eq}] and $c\IN(\phi,c_0,f)$ [\cref{eq:mean_cin}]. The color-coded lines [see colorbar in panel (c)] depict $c\IN(\phi,c_0,f)$ in the attractive membrane with $\mathcal{P}(\phi\C)=10 D_0$ (cf. \cref{fig:cin_of_phi_and_f}\pan{(c)}), for three different bulk concentrations $c_0$ as indicated above the panels. The black lines depict the (swollen-to-collapsed) transition function $\phi(c\IN)$ [\cref{eq:phi_eq}] for three different values of the transition sharpness $\Delta c$ [see legend in panel (a)]. Each interception point of one colored line and one black line refers to a steady-state solution ($c\IN^*,\phi^*$) that depends on $c_0$, $\Delta c$ and $f$. The solutions, $\phi^*(f)$, are summarized in \cref{fig:phi_steady_state}. In panel (a), we show $c_0=c_\tx{c}{D_\tx{in}(\phi_\tx{c})}/{D_0}$, i.e., the high-force limit (red dotted line) intercepts with the crossover point. In panel (c), we show $c_0=c_\tx{c}/{\mathcal{K}(\phi_\tx{c})}$, and the zero force limit (blue dotted line) intercepts with the crossover point $(c\C,\phi\C)$. In panel (b) the geometric mean of the two limits is chosen, i.e., $c_0=c_\tx{c}\sqrt{{D_\tx{in}(\phi_\tx{c})}/\left({D_0\mathcal{K}(\phi_\tx{c})}\right)}$. In this phase plane, $c_0$ performs a horizontal shift of $c\IN/c\C$. \label{fig:P10phaseplane}}
\ \\ \ \\
\includegraphics{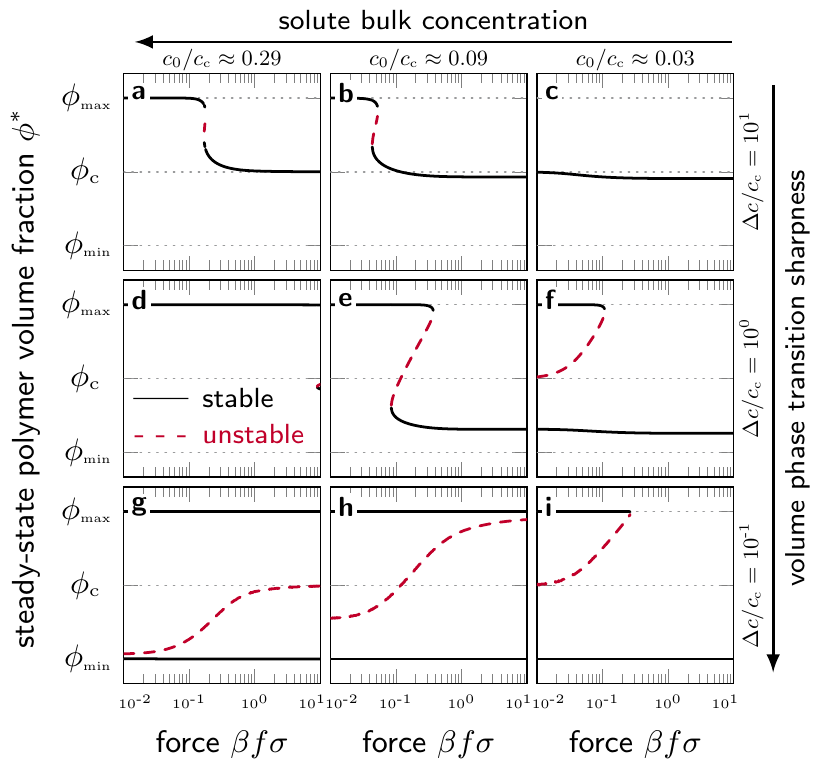}
\caption{\label{fig:phi_steady_state} Steady-state solutions of the polymer volume fraction, $\phi^*$, for attractive membranes ($\mathcal{P}(\phi\tb{c})=10D_0$) as function of the external driving force, $f$. The columns differ in terms of the bulk concentration, i.e., $c_0=c_\tx{c}{D_\tx{in}(\phi_\tx{c})}/{D_0}$ (panels in the left column), $c_0=c_\tx{c}\sqrt{{D_\tx{in}(\phi_\tx{c})}\left({D_0\mathcal{K}(\phi_\tx{c})}\right)}$ (central column), $c_0=c_\tx{c}/{\mathcal{K}(\phi_\tx{c})}$ (right column). Each row refers to one value of the transition sharpness, $\Delta c$ (see labels right of rows). In general, the force $f$ tunes $\phi^*$ from high ($\phi\tb{max}$) to low ($\phi\tb{min}$) values (since $c\IN(f)$ decreases for attractive membranes). One observes regions of multiple steady states (with two stable branches and one unstable branch) which may occur in the entire force range [e.g., panels (g) and (h)].} 
\end{figure*}

\subsection{Multiple steady-state solution}
With \cref{eq:phi_eq,eq:mean_cin} the feedback loop depicted in \cref{fig:mini_loop} is closed. We obtain numerically the steady-state solutions, $(c\tb{in}^*,\phi^*,)$, by finding the intersection points of $c\tb{in}(\phi,f)$  and $\phi(c\tb{in})$ in the phase plane. In this section we show the results with the attractive membrane only and demonstrate the general procedure. (For the repulsive membrane, we show a representative phase plane in \cref{fig:P01phaseplane} in Appendix \ref{sec:phaseplanerepulsive}.)

In \cref{fig:P10phaseplane}, the black lines depict the polymer's volume phase transition of $\phi(c\tb{in})$, \cref{eq:phi_eq}, for three different values of $\Delta c/c\tb{c}\in\left\{0.1,1,10 \right\}$. The colored lines, $c\tb{in}(c_0,f,\phi)$, \cref{eq:mean_cin},  are the inverted images of \cref{fig:cin_of_phi_and_f}\pan{(c)} and are plotted in \cref{fig:P10phaseplane} for three different bulk concentrations, $c_0$, from high [panel (a)] to low values [panel (c)]. As obvious, changing $c_0$ performs a shift of $c\IN(c_0,f,\phi)$ along the horizontal axis. In panel (a), we choose $c_0=c\tb{c}D\tb{in}(\phi_c)/D_0\approx0.29$ such that the high-force limit of $c\tb{in}$ intercepts with the crossover point $(c\tb{c},\phi\tb{c})$. In panel (c), we impose that the low-force limit intercepts with the crossover point, i.e., $c_0=c\tb{c}/(\mathcal{K}(\phi_c)\approx0.03$. In panel (b), the geometric mean, $c_0=c\tb{c}\sqrt{D\tb{in}(\phi_c)/(D_0\mathcal{K}(\phi\tb{c}))}\approx0.09$ is used as intermediate probe concentration.

For fixed $f$, $c_0$ and $\Delta c$, we find one or three interception points of $c\tb{in}(c_0f,\phi)$ and $\phi(c\tb{in})$, yielding the steady-state solutions, $\left(c\tb{in}^{*},\phi^*\right)$. In \cref{fig:P10phaseplane}\pan{(b)}, for instance, the low-force limit (blue dotted line) intercepts with the black solid line ($\Delta c/c\tb{c}=1$) only once at $\phi^*\approx\phi_\infty$, while it has three interceptions with the black dotted line ($\Delta c/c\tb{c}=0.1$) at $\phi_1^{*}\approx\phi\tb{min}$, $\phi_2^{*}\approx0.15$, and $\phi_3^{*}\approx\phi\tb{max}$.  In the case of triple solutions, the intermediate one is an unstable solution, while the other two are stable solutions. Precisely, the latter correspond to asymptotically stable solutions of the time-dependent Smoluchowski equation, $\dot c(z,t)=-\partial j(z)/\partial z$, i.e., the steady-state solution, $c^{*}(z)$, is restored after a small perturbation. A more detailed discussion on the stability of multiple solutions and consequences for the bistable domains is provided in a separate section below. 

We summarize the steady-state solutions for the attractive membrane by plotting $\phi^*(f)$ for different $c_0$, and $\Delta c$ in \cref{fig:phi_steady_state}. We observe a swelling (decrease in $\phi$) with hysteresis due to an increase in $f$ [see panels (a), (b), and (e)]. In more detail, higher $c_0$ can shift the force-induced $\phi$-transition to higher force values [e.g., compare panels (a) and (b)] and whether transitions may occur at all. For instance, as in the case of low transition sharpness, $\Delta c/c\tb{c}=10$, and low solute concentration [panel (c)], there is no significant effect on $\phi^*$. Similarly for the moderate sharpness, $\Delta c/c\tb{c}=1$, no transition is induced if $c_0$ is too high [panel (d)]. 

Further, $\Delta c$, plays an important role as it tunes the width of the bistable domains, e.g., while only small force ranges with bistability are observed for weakly responsive membranes ($\Delta c/c\tb{c}=10$), see \Cref{fig:phi_steady_state}\pan{(a)} and \labelcref{fig:phi_steady_state}\pan{(b)}, it can exist in the entire force range for sufficiently sharp transitions ($\Delta c/c\tb{c}=0.1$), see \Cref{fig:phi_steady_state}\pan{(g)} and \labelcref{fig:phi_steady_state}\pan{(h)}. 

We already conclude that the membrane's feedback response can lead to large bistable domains in $\phi$ tuned by $f$ and $c_0$, which is characterized by drastic switching of the membrane properties, such as the permeability, due to the bifurcations at the critical values.

\begin{figure*}\centering
\begin{minipage}{0.49\textwidth}\centering
\bf\sffamily attractive membrane\\ %$\mathcal{P}\tb{mem}(\phi\tb{c})=10^1D_0$ \\
\includegraphics{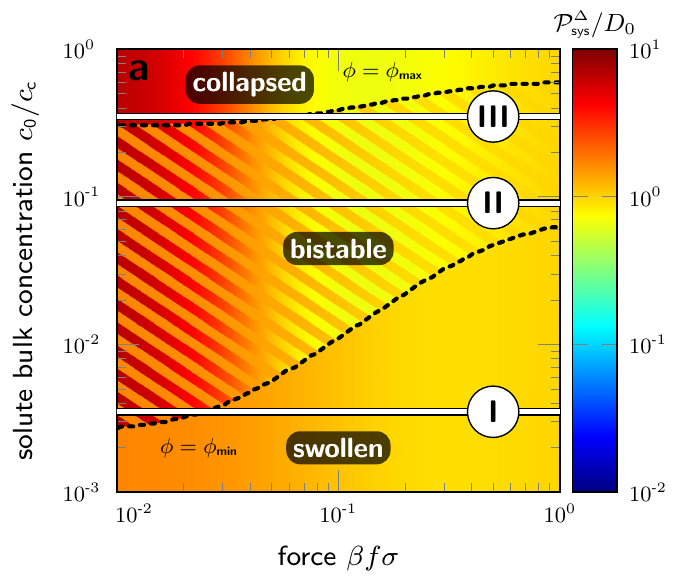}
\ \\
\includegraphics{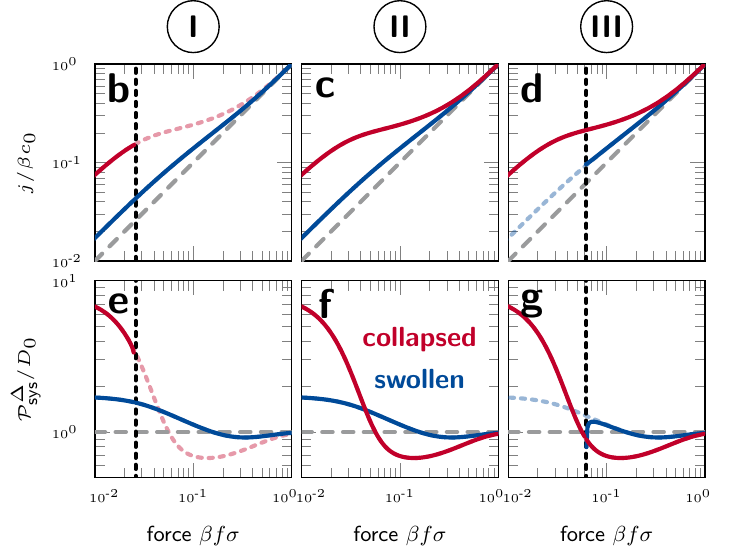}\\
\end{minipage}\vline
\begin{minipage}{0.49\textwidth}\centering
\bf\sffamily repulsive membrane\\ %$\mathcal{P}\tb{mem}(\phi\tb{c})=10^1D_0$ \\
\includegraphics{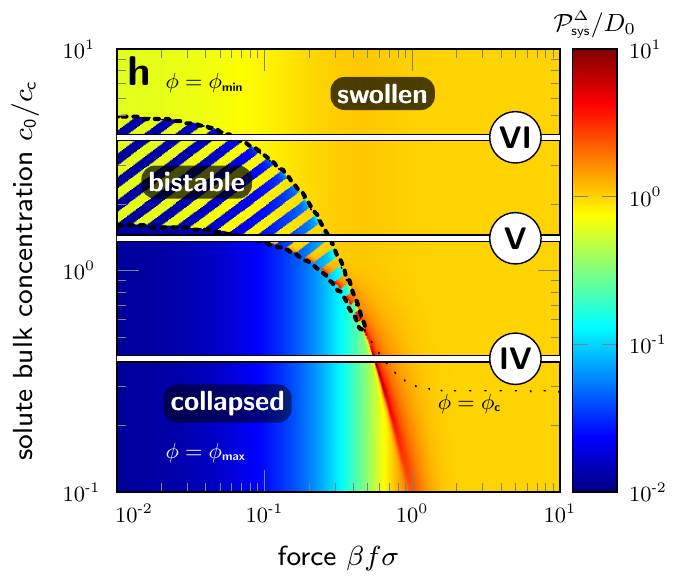}\\
\ \\
\includegraphics{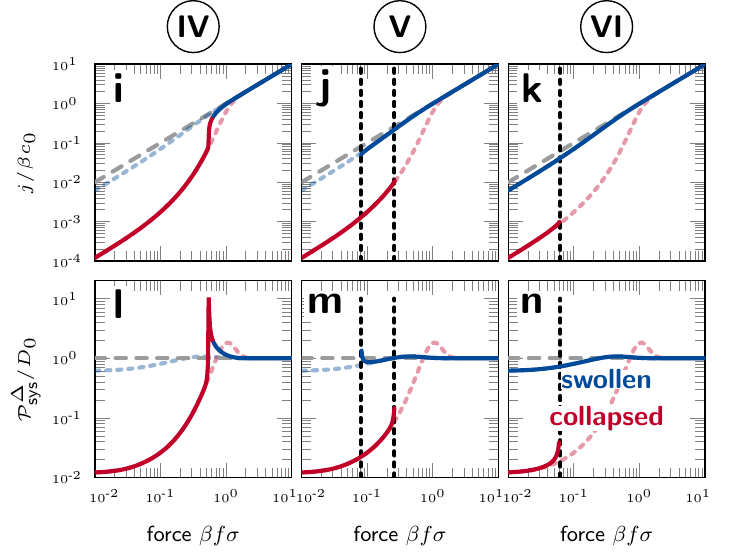}\\
\end{minipage}

\caption{\label{fig:Psys} Force-dependent permeability and flux of responsive membranes undergoing a very sharp volume phase transition with $\Delta c=0.1 c\C$. All panels on the left (right) hand side show the results for the attractive (repulsive) membrane. Top panels (a) and (h) depict the system's differential permeability, $\mathcal{P}\tb{sys}^{\Delta}/D_0$, as heatmaps in the $f$-$c_0$ plane. The heatmaps share the same color-code ranging from $10^{-2}$ to $10^{1}$ (see colorbar). The white lines labeled with roman numbers, I-VI, depict selected values of $c_0$, for which $j$ and $\mathcal{P}\tb{sys}^{\Delta}$ are presented in the panels below. The black dotted lines indicate the bifurcation at which the system changes from mono- to bi-stable (or vice versa), while the two solutions in the bistable domain are presented in a striped pattern.  In examples I-VI, the solutions are distinguished by the membrane's volume phase, i.e., blue corresponds to $\phi<\phi\tb{min}$ (swollen), and red to $\phi>\phi\tb{max}$ (collapsed). In fact, we find $\phi$ is either fully swollen or collapsed except for example IV, where a gradual crossover from $\phi(f=0)=\phi\tb{max}$ to $\phi(f\to\infty)\approx0.15$ is observed [cf. panel (h),  where the loosely dotted line indicates $\phi=\phi\C$]. The pale red and blue dotted lines in I-VI are the references for nonresponsive membranes in the fully collapsed and swollen case, respectively. The gray dashed lines in I-VI correspond to the bulk references, i.e., $j=D_0 c_0 \beta f $ and $\mathcal{P}\tb{sys}=D_0$, respectively, which yield the asymptotic values for $f\to\infty$ and $\phi\to0$. More details are provided in the main text.
}
\end{figure*}

\subsection{Consequences for the transport properties} 
The flux $j(f)$, given by \cref{eq:j_general_easy}, is a nonlinear function of $f$ determined by two contributions: The change in the membrane permeability $\mathcal{P}\tb{mem}(\phi)$ (due to the change in $\phi$) and the spatial setup (see Appendix \ref{sec:forcelimit} and our previous work\cite{Kim2022}). The nonlinear characteristics of $j(f)$ are quantified by the \emph{differential} system permeability,\cite{Kim2022} defined as
\EQ{
\mathcal{P}\tu{$\Delta$}\tb{sys}(f)=\frac{1}{\beta c_0}\frac{\mathrm{d} j}{\mathrm{d} f},\label{eq:Psys}
}
which describes the change in the steady-state flux induced by a change in the external driving force. 

We make use of $\mathcal{P}\tu{$\Delta$}\tb{sys}(f)$ to highlight the novel nonlinear effects on $j(f)$ caused by the membrane's feedback response. We limit ourselves to the very sharp membrane response ($\Delta c/c\tb{c}=0.1$), and point out the significant difference between the fluxes in attractive and repulsive membranes.

In \cref{fig:Psys}, we present heatmaps of $\mathcal{P}\tu{$\Delta$}\tb{sys}$ in the $f$-$c_0$ plane for the attractive [panel (a)] and the repulsive membrane [panel (h)]. The white lines labeled with roman numerals depict selected values of $c_0$, and correspond to the panels below, showing $j$ and $\mathcal{P}\tu{$\Delta$}\tb{sys}$. The heatmaps share the same color-scale (see colorbars), allowing a direct comparison between the results of attractive and repulsive membranes.

The attractive membrane exhibits, in general, larger $\mathcal{P}\tu{$\Delta$}\tb{sys}$ values than the repulsive one, particularly in the low-force and collapsed regime ($\phi=\phi\tb{max}$), in which the influence of $\mathcal{P}\tb{mem}$ is the greatest. For $f\to0$, we find $\mathcal{P}\tu{$\Delta$}\tb{sys}\approx 7 D_0$ and $\mathcal{P}\tu{$\Delta$}\tb{sys}\approx0.01 D_0$ for the attractive and repulsive membrane in the collapsed state, respectively. In the high-force limit, the system permeability converges to $D_0$ irrespective of the volume phase. If the membrane is swollen ($\phi=\phi\tb{min}$), the permeability of the repulsive and the attractive membrane are of the same order of magnitude, i.e., $\mathcal{P}\tb{mem}(\phi\tb{min})\approx 0.6 D_0$ and $\mathcal{P}\tb{mem}(\phi\tb{min})\approx 2 D_0$, respectively, and $\mathcal{P}\tu{$\Delta$}\tb{sys}$ does not deviate significantly from bulk diffusivity $D_0$, even for low forces (compare blue lines in lower panels of \cref{fig:Psys}).

Due to the sharp response with $\Delta c = 0.1 c_c$, the membrane is either fully swollen ($\phi\tb{min}$), or fully collapsed ($\phi\tb{max}$). 
Moreover, this also leads to large bistable domains in the $c_0$-$f$ plane, visualized as striped patterns in \Cref{fig:Psys}\pan{(a)} and \labelcref{fig:Psys}\pan{(h)}.
Crossing the boundary of these domains leads to a discontinuous volume phase transitions accompanied with an order of magnitude change in the solute flux (examples I, III, V, and VI). In the case of the repulsive membrane, the flux can be switched even by two orders of magnitude, particularly for small $f$.

In examples I-VI [\Cref{fig:Psys}\pan{(b)}--\labelcref{fig:Psys}\pan{(g)}, and \labelcref{fig:Psys}\pan{(i)}--\labelcref{fig:Psys}\pan{(n)}], we also depict the results for nonresponsive membranes in the fully swollen and collapsed case for comparison. In the nonresponsive case, $\mathcal{P}\tu{$\Delta$}\tb{sys}$ can also be tuned in the same range by only controlling $f$. Nonetheless, the membrane's responsiveness brings about more dramatic effects, such as bistability (I, III, V, VI) and hysteresis (V), yielding new control mechanisms to switch between two flux states. While nonresponsive membranes require large forces to exhibit bulk-like properties ($D_0$), a transition to this neutral state can be achieved in responsive membranes in a much sharper fashion and for lower force values, e.g., see \cref{fig:Psys} I, V, and VI. In V, for example, the crossover from the low- to the high-permeability state occurs abruptly around $\beta f \sigma\approx 0.2$ (red solid line), whereas for the collapsed, nonresponsive membrane a gradual change is observed in the range $\beta f \sigma\approx 0.1-1.0$.

Further, even if the polymer volume phase transition occurs without bifurcation, nonlinearities in the force--flux relations can be significantly enhanced. For instance, in panel (l) (example IV), we find a tenfold maximization of $\mathcal{P}\tu{$\Delta$}\tb{sys}\approx12D_0$ at roughly $\beta f\sigma\approx 0.5$, which even exceeds the maximum differential system permeability measured for the attractive membrane.

{\subsection{Discussion: nonequilibrium steady-state stability}

Our model results into well-defined force--flux relations in the domains with unique steady-state solutions. In the bistable domains, however, the question arises whether the states coexist or whether only one survives under real conditions. This far, the solutions were simply deduced from a deterministic interpretation of the macroscopic model equations, i.e., by evaluating  the self-consistency equation $c\IN=R(c\IN)$, with $R(c\IN)=d^{-1}\int\IN c(z,\phi(c\IN)) \mathrm{d}z$. The steady state is asymptotically stable, if ${\mathrm{d}R(c\IN)}/{\mathrm{d}c\IN}|_{c\IN^*}<1$.\cite{Frank2004} However, our approach neglects larger fluctuations in $\phi$ and $c\IN$, and does not analyze further nonequilibrium extremum principles.\cite{Donskoy2022} In the following, we first discuss the consequences of the deterministic perspective, and then briefly review alternative interpretations.

In the case of negligible fluctuations the (deterministic) transition between states can be either \emph{reversible} or \emph{irreversible}.\cite{Hahn1973} One \emph{reversible} transition is example V [\Cref{fig:Psys}\pan{(j)} and \labelcref{fig:Psys}\pan{(m)}]. Here, the membrane is in the collapsed state (red line) for small $f$. With increasing $f$, the membrane is driven into the bistable regime, yet remains in the collapsed state. Only if $f$ exceeds the second bifurcation line, the membrane swells. In the same example V, if $f$ is decreased from high force values, the membrane stays swollen in the bistable domain and returns to the collapsed state until the first bifurcation line is passed. Hence,  we find a \emph{reversible} transition with hysteresis between the two state in V.

In contrast, consider example I or VI, and assume that the membrane is in the collapsed state at $f=0$, an increase in $f$ leads to a swelling when the bifurcation line is crossed. Decreasing the force again, however, does not induce a collapse, and, hence, this transition can be termed \emph{irreversible} in the deterministic interpretation. This is because only the swollen case survives once the threshold is surpassed. Analogously, see example III,  a collapsed-to-swollen transition cannot be induced by increasing $f$.

Although two stable states may coexist in the deterministic model, 
one of them could be metastable and practically unoccupied under experimental conditions. In literature one finds nonequilibrium principles, e.g., based on the maximization of entropy, the minimization of entropy production (least dissipation), or the minimization of power, providing various possible routes.\cite{Martyushev2006,P.Glansdorff,Kondepudi1998,Attard2009,Katchalsky1968,Tome2006,Graham1985,Seifert2012,Donskoy2022, Schmidt2022} Such extremum principles may lead to unique solutions in the \emph{bistable} regime, and to different values for $f$ and $c_0$, where the switching between the high and low flux states occurs. For example, it should be the minimum flux, if the least-dissipation principle applies. This has direct consequences on the flux--force relation and the critical transition values of $f$ and $c_0$, where the phase transition in V would occur always at the first bifurcation line, i.e., without bistability and hysteresis.

Furthermore, the presented diffusion process can also be modeled with the stochastic Smoluchowski equation,\cite{Chavanis2008} and possibly further coarse-grained to a stochastic differential equation for $\dot c\IN$.\cite{Archer2004,Gardiner2004} Hence, given the fluctuations are large enough, a stochastic switching between the two steady states may be observed in the bistable domains, and the effective force--flux relations are determined by the averaged values of $\phi$ and $c\IN$. Consequently, changing $f$ results in a continuous transition between the two states, implying that example V does not exhibit hysteresis behavior, but is rather similar to the transition in example IV.

 Ultimately, the appropriate stability interpretation remains to be verified, and is likely specific to the membrane material and the experimental nonequilibrium conditions. Nonetheless, a strong amplification of nonlinear characteristics and a critical switching in the force--flux relations can be expected due to the membrane's responsiveness.

\section{Summary}
We have investigated the driven steady-state solute transport through polymeric membranes with a sigmoidal volume phase response to the penetrant uptake. The change in the polymer volume fraction is decisive for the membrane permeability, which we modeled with exponential functions. This, in turn, impacts on the solute uptake, leading to novel feedback-induced effects in force--flux relations that cannot be achieved by nonresponsive membranes. We quantified our findings in terms of the system's differential permeability. 

The feedback effects of responsive membranes are most pronounced in the low-force regime, where the bulk concentration largely tunes the membrane density between the swollen and the collapsed state. Increasing the force can lead to a membrane swelling accompanied with a strong amplification of nonlinear characteristics and critical switching in the force--flux relations. For instance, the swelling of membranes with repulsive polymer-solute interactions can be caused by a small change in the driving force, for which we report an increase in the flux by two orders of magnitude, and a pronounced maximization of the differential permeability, i.e., a tenfold increase compared to the case of nonresponsive membranes.

Moreover, of particular note is the feedback-induced coexistence of two stable steady states, while the size of the bistable domains increases with the sharpness of the sigmoidal polymer response.  The bifurcations from mono- to bistability occur at critical values of the driving force, and the solute bulk concentration, leading to discontinuous changes in the flux of up to two orders of magnitude. 

Thus, the force-dependent switching between high and low flux states provides a valuable control mechanism for molecular transport. It can be fine-tuned also to control the appearance of hysteresis, enabling the presented feedback membranes to function as memristive devices. Moreover, the coupling of the permeability hysteresis to (non-oscillatory) chemical reactions may lead to biomimetic features, such as membrane excitability and autonomous oscillations, as first proposed by theory,\cite{Hahn1973,Siegel1995, Katchalsky1968, Kepperbook} and eventually validated by experiments.\cite{Bell2021a,Zou1999,Leroux1999}

 Hysteresis transitions found in literature\cite{Bell2021a,Zou1999,Leroux1999,Katchalsky1972,Koga2001,Kamemaru2018,Annaka1992} are usually rationalized by a bistability in the polymer's conformational free energy,\cite{Hiller2003,Baker1996,Annaka1992} and attributed to the complex microscopic interactions or the competition between entropic and energetic contributions.\cite{Hiller2003,Baker1996,Annaka1992}  
 Nonetheless, many polymers exhibit a hysteresis-free response, for which the presented feedback mechanism provides a novel explanation of how hysteresis transitions  can be generated and tuned in polymer membranes.
 
We disclosed in this work how nonlinear solute transport through chemo-responsive polymer membranes is controlled by membrane feedback. It thus provides the theoretical basis for the rational design of self-regulating membranes with nonlinear control features for molecular transport. Adaptations employing more complex functions for partitioning, diffusivity, and the polymer response, to differing feed and permeate bulk concentrations as well as extensions to different spatial arrangements and geometries could be interesting for future studies.

\section*{Acknowledgments}
The authors thank Matej Kandu\v{c} for fruitful discussions. This work was supported by the Deutsche Forschungsgemeinschaft (DFG) via the Research Unit FOR 5099 ``Reducing complexity of nonequilibrium systems''. W.K.K. acknowledges the support by the KIAS Individual Grants (CG076001 and CG076002) at Korea Institute for Advanced Study. 

\section*{Author declarations}

\subsection*{Conflict of Interest}
The authors have no conflicts to disclose.

\appendix

\section{The flux in the low- and high-force limits \label{sec:forcelimit}}
The flux, $j$ [\cref{eq:j_general_easy}], in the small-force regime reads\cite{Kim2022}
\EQ{\lim_{f\to0}j=D_0 \beta f c_0\left[1+\left(\frac{D_0}{\mathcal{P}\tb{mem}}-1\right)\frac{d}{L}\right]^{-1}, \label{eq:Psys_low_f}
}
which converges to $\lim_{f\to0}j\to\mathcal{P}\tb{mem}c_0 \beta f$ for $d\to L$. The membrane thickness, $d$, determines the crossover to the high-force regime, for which $\lim_{f\to\infty}j=D_0c_0\beta f$ results. 

For moderate to large forces, we find $S(f)\approx \exp(-\beta f (L-d)/2)$. With increasing $f$, the denominator in \cref{eq:j_general_easy} converges to unity, governed by $(L-d)$. So, the larger $d$ with respect to $L$, the higher $f$ has to be in order to reach the high-force limit. Obviously, the onset of the high-force limit also depends on the membrane permeability, precisely, large values of $\mathcal{P}\tb{mem}$ will shift the crossover to smaller force values.

\section{Concentration profiles \label{sec:profile}}
The solute concentration profile in a system depicted in described in \cref{sec:smolu} reads\cite{Kim2022}
\EQ{
c(z)=c_0 \left[1-\dfrac{D_0\beta f \mathcal{I}(0,z)}{1+\left(\frac{D_0}{\mathcal{P}\tb{mem}}-1\right)S(f)}\right]e^{-\beta\left(G(z)-fz\right)}, \label{eq:cinofz}
}
We can split \cref{eq:cinofz} into the piecewise homogeneous layers, precisely, the feed boundary, membrane (`in'), and permeate boundary layer, and can write the respective full expressions as
\EQ{&c(z)|\tb{feed}=\nonumber\\
  &c_0\mathcal{K}\ \dfrac{e^{\beta f z} \left({D_0}-\mathcal{P}\tb{mem}\right) S(f)+{P\tb{mem}} }{({D_0}-\mathcal{P}\tb{mem}) S(f)+\mathcal{P}\tb{mem} },\label{eq:cfeed} \\ \nonumber \\
&c(z)|\IN=\nonumber\\
  &c_0\mathcal{K}\ \dfrac{\frac{e^{\beta f \left(z-L/2\right)}}{\sinh(\beta f L/2)}\left({D_0}-\mathcal{P}\tb{mem}\right) {\sinh \left(\beta f \frac{d-L}{2}\right)}+{D_0} }{({D_0}-\mathcal{P}\tb{mem}) S(f)+\mathcal{P}\tb{mem} },\label{eq:cin} \\ \nonumber \\
  &c(z)|\tb{permeate}=\nonumber\\
  &c_0\mathcal{K}\ \dfrac{e^{\beta f (z-L)} \left({D_0}-\mathcal{P}\tb{mem}\right) S(f)+{P\tb{mem}} }{({D_0}-\mathcal{P}\tb{mem}) S(f)+\mathcal{P}\tb{mem}}.\label{eq:cperm}
}

 \section{Phase plane for repulsive membranes\label{sec:phaseplanerepulsive}}
\begin{figure}[h!]
\includegraphics{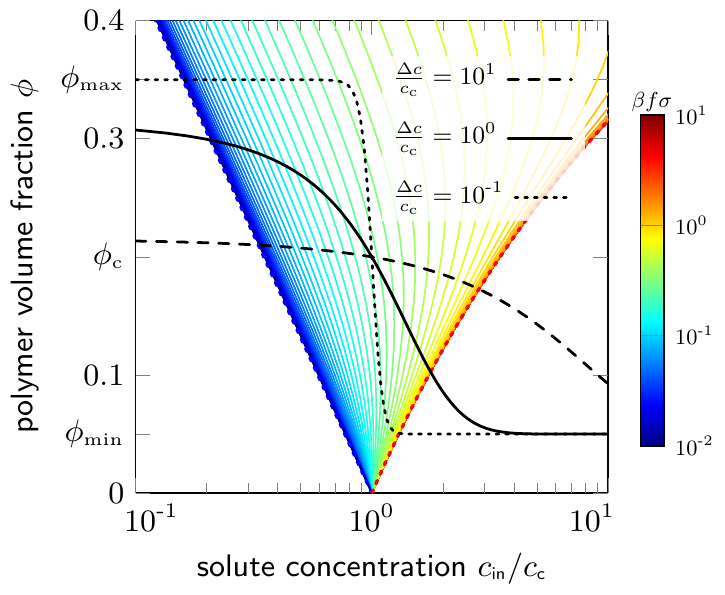}
\caption{Phase plane showing
$\phi(c\IN)$ [\cref{eq:phi_eq}] and $c\IN(\phi,c_0,f)$ [\cref{eq:mean_cin}]. The color-coded lines (see colorbar) depict $c\IN(\phi,c_0,f)$ for the repulsive membrane ($\mathcal{P}(\phi\C)=0.1 D_0$, cf. \cref{fig:cin_of_phi_and_f}\pan{(a)}), with selected probe concentration $c_0$. The black lines depict the (collapsed-to-swollen) transition function $\phi(c\IN)$ [\cref{eq:phi_eq}] for three different values of the transition sharpness $\Delta c$ (see legend). Each interception point of a colored line [\cref{eq:mean_cin}] and a black line [\cref{eq:phi_eq}] refers to a steady-state solution ($c\IN^*,\phi^*$) that depends on $c_0$, $\Delta c$ and $f$.  \label{fig:P01phaseplane}}
\end{figure}

 In \cref{fig:P01phaseplane}, $c\IN(c_0,f,\phi)$ [\cref{eq:mean_cin}] and  $\phi(c\IN)$ [\cref{eq:phi_eq}] are presented in the $c_0$-$\phi$ plane. Interception points of $c\IN$ and $\phi$  correspond to the force-dependent steady-state solution. The results were used to calculate $\phi^*(c_0,f)$, which enter the flux and the differential permeability as depicted in \Cref{fig:Psys}\pan{(h)}--\pan{(n)}. The bulk concentration, $c_0$, and the external force, $f$, yield, in general, an increase in the mean inside concentration, $c\IN$, resulting in a swelling of the membrane. For very sharp transitions, e.g., $\Delta c= 0.1c\C$, one can find multiple solutions for fixed $c_0$ and $f$, giving rise to bistability and hysteresis.

\end{document}